\documentstyle[aps,pre,graphicx,multicol]{revtex}
\begin{document}
\draft

\title{Dynamic behavior of driven interfaces 
       in models with two absorbing states}

\author{Sungchul Kwon$^1$, WonMuk Hwang$^2$ and Hyunggyu Park$^1$}
\address{$^1$ Department of Physics, Inha University,
             Inchon 402-751, Korea}
\address{$^2$ Department of Physics, Boston University,
             Boston, MA 02215, U.S.A.}

\date{\today}
\maketitle

\begin{abstract}

We study the dynamics of an interface (active domain) between different absorbing 
regions in models with two absorbing states in one dimension;
probabilistic cellular automata models and interacting monomer-dimer models. 
These models exhibit 
a continuous transition from an active phase into an absorbing phase, which
belongs to the directed Ising (DI) universality class. 
In the active phase, the interface spreads ballistically into the absorbing regions 
and the interface width diverges linearly in time. 
Approaching the critical point, the spreading velocity of the interface vanishes 
algebraically with a DI critical exponent.
Introducing a symmetry-breaking field $h$ that prefers one absorbing state over
the other drives the interface to move asymmetrically  toward
the unpreferred absorbing region.
In Monte Carlo simulations, we find that the spreading velocity of this driven interface 
shows a discontinuous jump at criticality. 
We explain that this unusual behavior is due to a finite relaxation time in the
absorbing phase. The crossover behavior from the symmetric case (DI class)
to the asymmetric case (directed percolation class) is also studied.                           
We find the scaling dimension of the symmetry-breaking field
$y_h = 1.21(5)$.
\end{abstract}


\begin{multicols}{2}
\narrowtext

Universality classes of models exhibiting a continuous phase transition
from an active phase into an absorbing phase with multiple absorbing states
are determined by the symmetry 
between the absorbing states
\cite{Park-95d,Park-98,Menyhard-96,Bassler-96,Hinrichsen-97}. 
If the absorbing phase consists of two equivalent absorbing states,
the phase transition belongs to the directed Ising (DI)
universality class
\cite{Park-98,Menyhard-96,Bassler-96,Hinrichsen-97,Grass-Krause-84,Park-94,Menyhard-94,Bassler-97a}. 
When the symmetry between the two absorbing
states is broken, one absorbing state becomes completely obsolete 
and the system crosses over to the directed percolation (DP) universality class
\cite{Park-95d,Park-98,Menyhard-96,Bassler-96,Hinrichsen-97}.
This crossover has been observed in many models with two absorbing states,
including the probabilistic cellular automata (PCA) model \cite{Grass-Krause-84} 
and the interacting
monomer-dimer (IMD) model \cite{Park-94} in one dimension.

Dynamic properties of an active domain in the sea of the absorbing region
are well known for models with a single
absorbing state, which belong to the DP universality class
\cite{Grass-Torre-79}.
The size of the active domain $R$ in one dimension is defined as the distance
between the two farthest active sites averaged over surviving samples and
satisfies the dynamic scaling relation 
\begin{equation}
R(t,\Delta) = t^{1/z} f(\Delta t^{1/\nu_\|}),
\label{eq-1}
\end{equation}
where 
$t$ is the time, $\Delta$ the reduced external parameter,
$z=\nu_\| /\nu_\bot$ the dynamic exponent \cite{explain2},
and $\nu_\|$ ($\nu_\bot$) the relaxation time (correlation length)
exponent.  

In the active phase ($\Delta >0$), the active domain 
spreads ballistically with a finite velocity and its size diverges
linearly in time. So $f(x)$ must scale as $x^{\nu_\| - \nu_\bot}$
in the $x\rightarrow +\infty$ limit. Therefore the
asymptotic value of the spreading velocity,
$v = \lim_{t\rightarrow\infty} R/t$, scales as 
\begin{equation}
\label{eq-2}
v(\Delta)\sim \Delta^{\nu_\| - \nu_\bot}.
\end{equation}
At criticality ($\Delta =0$), the active domain spreads 
algebraically in time,
$R(t,0) \sim t^{1/z}$ with the scaling function $f(0)$
being a constant.
In the absorbing phase ($\Delta <0$), 
the active domain diffuses like a random walker but its size 
converges to a finite value of order of the 
correlation length. So $f(x)$ must scale as $(-x)^{- \nu_\bot}$
in the $x\rightarrow -\infty$ limit. Therefore 
the size of the active domain in the long time limit scales
as 
\begin{equation}
\label{eq-3}
R(\infty,\Delta)\sim (-\Delta)^{-\nu_\bot},
\end{equation}
and the spreading velocity $v=0$ for $\Delta \le 0$.

For models with two absorbing states, one can consider 
the dynamics of an active domain in two different environments,
i.e., in the sea of one absorbing region (defect dynamics) and
between two different absorbing regions (interface dynamics)
\cite{Park-98,Park-94}.
The defect dynamics describe the spreading of a defect (active domain)
in a nearly absorbing space, while the interface dynamics
describe the spreading of an interface (active domain) between
two different absorbing regions. 
Note that the interface can not disappear by itself
in contrast to the defect.
           
For the symmetric case, the system belongs to the DI universality class
and the active domain (defect or interface) behaves similarly in 
both dynamics.
In the active phase, the spreading velocity
scales as in Eq.\ (\ref{eq-2}) with the DI critical exponents.
At criticality, the size of the active domain grows as
$R(t,0) \sim t^{1/z}$ with the DI dynamic exponent $z$.
In the absorbing phase, 
the active domain splits into two branches, which
diffuse like two random walkers before they meet and
annihilate each other.
The absorbing domain of the different type from the absorbing sea
emerges between the two 
active branches. The width of each branch should scale as in 
Eq.\ (\ref{eq-3}) with the DI exponent but the 
size of the active domain defined as the distance 
between the two farthest active sites must be of order of
the distance between two random walkers. So it
diverges with the random walk dynamic exponent, $R \sim t^{1/z_{RW}}$
with $z_{RW} = 2$.

With a symmetry-breaking field that prefers one absorbing state over
the other, the system crosses over to the DP universality class.
The active domain (defect) in the sea of the preferred absorbing
region behaves in the exactly same way as in the defect dynamics of 
models with a single absorbing state, because the probability of
creating the unpreferred absorbing domain is exponentially small. 
The spreading velocity in the active phase 
scales as in Eq.\ (\ref{eq-2}) with the DP exponents and
the defect size in the absorbing phase as in Eq.\ (\ref{eq-3}) 
with the DP exponent. 

However, the dynamics of
the active domain (interface) between the preferred and the unpreferred absorbing
region are completely different.
The interface is now driven into the unpreferred absorbing region by the
symmetry-breaking field.
In the active phase, 
the interface spreads ballistically in both directions.
Of course, the interface front near the preferred absorbing region ($P$-interface front)
moves slower than the interface front near the unpreferred absorbing
region ($U$-interface front). The unpreferred absorbing state is not responsible
for the absorbing phase transition and 
its region always tends to shrink against the preferred absorbing region.
So the dynamic behavior of the $U$-interface front is always ballistic
even in the absorbing phase and at criticality. 
The $U$-interface front velocity, $v_U$, varies smoothly 
with the external parameter $\Delta$ and
there is no singularity at the critical point. 
The $P$-interface front behaves like a boundary of the active 
domain in the sea of the preferred absorbing region (defect dynamics).
The $P$-interface front velocity, $v_P$, scales as in Eq.\ (\ref{eq-2}) with
the DP exponents in the active phase.
At criticality, the $P$-interface front behaves in a critical fashion of the DP type.
The average spreading distance of the $P$-interface front from its initial position 
scales algebraically as $t^{1/z}$ with the DP dynamic exponent $z$~\cite{Park-98}.                                                                 
However, the size of the active domain (interface width) is the average 
distance between the $U$-interface front and the $P$-interface front, so diverges
still ballistically even at criticality, $R(t,0) \simeq v_U t$.

As we cross the critical point into the absorbing phase, an interesting thing
happens. In the absorbing phase, 
the interface still grows ballistically into the unpreferred absorbing region
with a finite value of $v_U$, but the preferred absorbing state 
dominates the system and tries to confine the interface within a finite region
like the active domain in the DP systems.
We expect that  the $P$-interface front moves asymptotically
with the same velocity and the same direction as the $U$-interface front
in the entire absorbing phase. 
Then $v_P$ must have a finite jump at the absorbing phase transition and
the interface width be finite in the long time limit.

In this paper, we study numerically the dynamics of these driven interfaces
in the PCA model and the IMD model with a symmetry-breaking field. 
We perform dynamic Monte Carlo simulations to measure the 
interface width and the spreading velocities of the two interface fronts.
Indeed, we find a discontinuous jump of $v_P$ at criticality and
an interesting scaling behavior of the interface width near the transition.

The PCA model studied here was originally introduced by
Grassberger, Krause, and von der Twer \cite{Grass-Krause-84}. The model evolves with
elementary rule number 94 in the notation of Wolfram
\cite{Wolfram-86} except $110$ and $011$ configurations, where the central spin 1 
flips to 0 with probability $p$ and remains unflipped with probability $1-p$.
This model has two equivalent absorbing states, i.e., ($1010\cdots$) and 
($0101\cdots$), and exhibits an absorbing phase transition
that belongs to the DI universality class.
We introduce a symmetry-breaking field $h$ that prefers           
($1010\cdots$) over ($0101\cdots$) \cite{Park-98}.
With probability $h$, we reject the flipping attempt 
of the central spin in the $111$ configuration when it is at an 
odd-numbered site. As the system must go through a $111$ configuration 
right before entering into an absorbing state, i.e., 
$(\cdots 010\underline{111}010\cdots)\rightarrow
(\cdots 010\underline{101}010\cdots)$, the absorbing state 
with 1's at the odd-numbered sites is 
probabilistically preferable to the other for nonzero $h$.

For the interface dynamics, we start with a single kink (domain wall)
between the two different absorbing regions, i.e.,
($\cdots 101\underline{00}101\cdots$), where 
the sharp interface (active domain) with zero width is placed in the middle of
$\underline{00}$. Then the system is updated in parallel following the 
automata rule. In contrast to the defect dynamics, the system never enters an 
absorbing state. We measure the positions of the $P$-interface front and
the $U$-interface front, averaged over $2\times 10^3$ samples up to typically 
$10^5$ time steps for various values of $p$ and $h$.               

Fig.\ 1 shows typical evolutions of the interface for (a)
$\Delta>0$, (b)  $\Delta=0$, and (c) $\Delta <0$ where
$\Delta \equiv p -p_c$ and the critical probability $p_c=0.3908(5)$
for $h=0.5$~\cite{explain0}. Evolutions of the average positions of the $P$- and $U$-interface fronts
are shown in (d). The $U$-interface front always moves ballistically
and its velocity $v_U$ changes smoothly with $\Delta$.
However, the $P$-interface front shows an abrupt change in its
dynamics at criticality. It moves ballistically in the active phase and
its velocity $v_P$ vanishes algebraically at criticality. 
The $P$-interface front 
still moves toward the preferred absorbing region 
at criticality and its average distance from the initial position
diverges as $t^{1/z}$ with the DP dynamic exponent $z$. 
In the absorbing phase, the $P$-interface front 
turns around and moves 
ballistically in the opposite direction. 
In the long time limit, the $P$-interface front moves with
the same velocity and the same direction as the $U$-interface front
for $\Delta <0$. 

In Fig.\ 2, we plot $v_P$ and $v_U$ versus $\Delta$ for $h=0$ and 
0.5. The critical probability $p_c = 0.125(2)$ for the symmetric case
($h=0$)~\cite{explain0}. Indeed, we find that there is a discontinuous jump of $v_P$ at criticality
for the asymmetric case ($h=0.5$) and  both velocities coincide in the absorbing phase.
In the active phase, $v_P$ vanishes continuously approaching the critical
point like in the symmetric case.
Log-log plots of $v_P$ versus $\Delta$ near criticality in the active phase 
are shown in Fig.\ 3.
From the slopes, we estimate $\nu_\| - \nu_\bot$ as 1.35(10) for $h=0$
and 0.65(3) for $h=0.5$, which agree well with the DI (1.39) and DP 
(0.637) value, respectively. 
We also run dynamic Monte Carlo simulations for the IMD model
and find the similar results, i.e., 1.25(15) for the symmetric case and
0.62(3) for the asymmetric case. 

In the absorbing phase, the average distance between the $P$-interface front
and the $U$-interface front (interface width) is found to saturate to a finite value, 
$R_s = R(\infty,\Delta)$. However, $R_s$ does not
scale as in Eq.(\ref{eq-3}) because this length scale 
is not proportional to the correlation length of the system.
As one can see in Fig.\ 1(c), this length does not
measure the typical size of the DP active domain but the distance
from the unpreferred absorbing region to the farthest active site
in the preferred absorbing region. 
As the $U$-interface front always moves ballistically into the unpreferred
absorbing region, this distance should be proportional to
the time scale in the system which measures the {\em average} lifetime of the
tree-like active DP clusters grown out of the $U$-interface front. However,
small clusters do not contribute effectively to the average lifetime
due to the {\em shading effect} by big clusters. This effect complicates
the scaling behavior of $R_s$.                    

A simple-minded scaling theory for $R(t,\Delta)$ in the asymmetric
interface dynamics may be given as
\begin{equation}
\label{eq-4}
R(t,\Delta) = t g(\Delta t^{1/\nu_\|}),
\end{equation}
where $\nu_\|$ is the relaxation time exponent for the DP universality class.
At $\Delta=0$, $g(0)$ is a constant, so $R(t,0)$ diverges linearly in time,
which is correct for the asymmetric case. For $\Delta <0$, 
$R(t,\Delta)$ should saturate in the long time limit,
so $g(x)$ must scale as $(-x)^{-\nu_\|}$ in the $x\rightarrow -\infty$
limit. Therefore we expect that $R_s$ scales as $(-\Delta)^{-\nu_\|}$
with the DP exponent $\nu_\| \simeq 1.733$. 
The log-log plot of $R_s$ versus $|\Delta|$ shows a fairly good straight line,
which seems to support a simple power-law scaling of $R_s$
(see Fig.\ 4).
However, our estimate for the scaling exponent from its slope
is well over the above DP value, i.e., 2.00(5) for the PCA model 
and 1.95(10) for the IMD model. 
We find that this discrepency is due to the shading effect.
A careful analysis incorporating the shading effect suggests
that there should be a logarithmic correction in the scaling theory,
i.e., $R_s \sim |\Delta|^{-\nu_\|} \ln |\Delta|$ \cite{Park-99e}. 
The log-log plot of $R_s /|\ln |\Delta||$ versus $|\Delta|$
(Fig.\ 4)
also shows a fairly good straight line.
From the slope, 
we estimate the scaling exponent $\nu_\|$ = 1.75(5) for
the PCA model and 1.65(10) for the IMD model, which
are in good agreement with the DP value.

Finally, we study the crossover behavior from the DI class to
the DP class. The operator associated with the
symmetry-breaking field $h$ must be relevant at the DI critical
point, so the scaling dimension of this crossover operator 
$y_h$ must be positive and may be an independent DI critical
exponent. We obtain $y_h$ numerically by measuring the
interface front velocities at small values of
$h$ along the $p=p_c^0$ line, where $p_c^0=0.125$ is the DI critical
probability at $h=0$. 
At finite values of $h$ along the $p_c=p_c^0$ line, the system becomes absorbing and
the two interface front velocities, $v_U$ and $v_P$,
coincide and become finite.

Consider the crossover scaling relation
near the DI critical point 
for the average position of the unfavorred interface front $X_U$ 
measured from its initial position as
$X_U (t,h) =b X_U (b^{-z_{DI}} t, b^{y_h} h)$ where
$z_{DI}$ is the DI dynamic exponent (1.75) and $b$ an arbitrary 
scaling factor. With $b=t^{1/z_{DI}}$, we have
$X_U=t^{1/z_{DI}}\Phi (h t^{y_h/z_{DI}})$.
For the symmetric case ($h=0$), 
$X_U$ scales as $t^{1/z_{DI}}$ with $\Phi(0)$ being a constant.
For $h > 0$, $X_U$ increases linearly in time (finite $v_U$), so
$\Phi(x) \sim x^{(z_{DI}-1)/y_h}$ in the $x\rightarrow \infty$ 
limit. Therefore, $v_U \sim h^\kappa$ where $\kappa= (z_{DI}-1)/y_h$.

Log-log plots of $v_U$ versus $h$ at $p=p_c^0$ are shown in Fig.\ 5.
From the slope, we estimate $\kappa =0.62(2)$ and hence obtain
$y_h = 1.21(5)$. Interestingly, this value is very close to the value of
the scaling dimension of the operator associated with a roughening
degree of freedom at the DI critical point~\cite{Park-99a}.
The conventional definition of the crossover exponent 
$\phi$ is given as the ratio of the two scaling dimensions, i.e.,
$\phi \equiv y_h/y_\Delta$ where $y_\Delta$ is the DI scaling dimension
of the temperature-like operator $\Delta$ and $y_\Delta = 1/\nu_\bot$.
Using this relation, we find $\phi=2.24(10)$ which is consistent with the recent result
for generalized monomer-monomer models studied by Bassler and 
Browne~\cite{Bassler-96}.

In summary,
we study numerically the dynamics of the driven interfaces in the PCA and IMD model 
that have two asymmetric absorbing states.
We find that the spreading velocity of the driven interface shows a discontinuous jump
at criticality due to the finite interface width in the absorbing phase.
The interface width diverges in a nontrivial manner, 
approaching the criticality.  We find that our
numerical data are consistent with a recent scaling theory taking into account 
the shading effect of big active clusters over small ones~\cite{Park-99e}.

We wish to thank C.-C. Chen and M. den Nijs for useful discussions.
This work was supported in part by the KOSEF through the
SRC program of SNU-CTP, by the Ministry of Education, Korea (97-2409),
and by the Inha University research grant (1997).

\begin{figure}
\centering
\includegraphics[width=8cm]{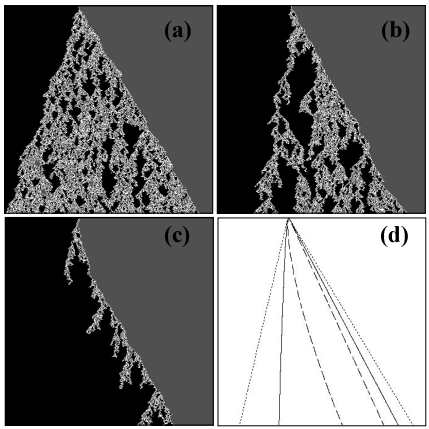}
\caption{
Typical evolutions of the asymmetric interface dynamics for (a) $\Delta=0.05$, (b) 
$\Delta = 0$, and (c) $\Delta=-0.05$ with $h=0.5$. The preferred (unpreferred) 
absorbing region is shown in black (grey) and the active sites are represented by
white pixels. Evolutions of the interface fronts averaged over many samples are
shown in (d): dotted lines for the active phase, solid lines at criticality,
and dashed lines for the absorbing phase.
}
\label{fig1}
\end{figure}

\begin{figure}
\centering
\includegraphics[height=8cm,angle=270]{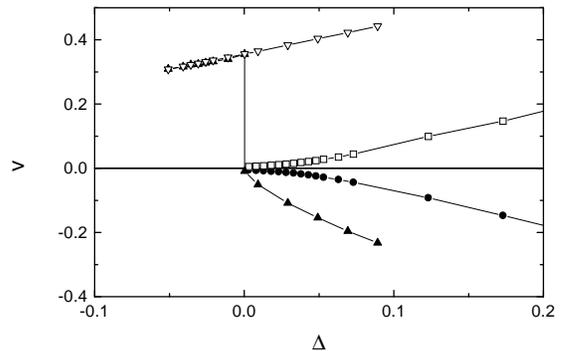}
\caption{
Interface front velocities versus $\Delta$. 
Filled circles and open boxes represent the interface front velocities for
the symmetric case. Filled up-triangles represent the $P$-interface front velocity
$v_P$ and open down-triangles the $U$-interface front velocity $v_U$ for
the asymmetric case ($h=0.5$). Lines between data points are guides to the eye.
}
\label{fig2}
\end{figure}

\begin{figure}
\centering
\includegraphics[height=8cm,angle=270]{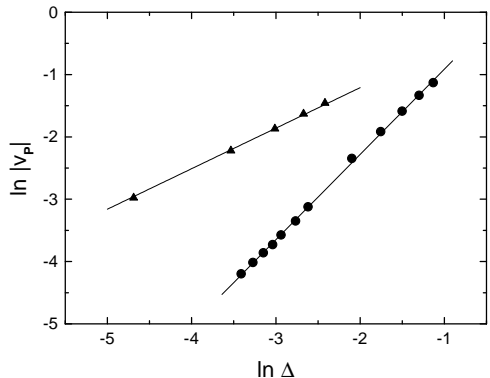}
\caption{
Log-log plots of $v_P$ versus $\Delta$. Filled circles are for the symmetric case
and filled triangles for the asymmetric case ($h=0.5$). The solid lines
are of slope 1.35 and 0.65, respectively.
}
\label{fig3}
\end{figure}

\begin{figure}
\centering
\includegraphics[height=8cm,angle=270]{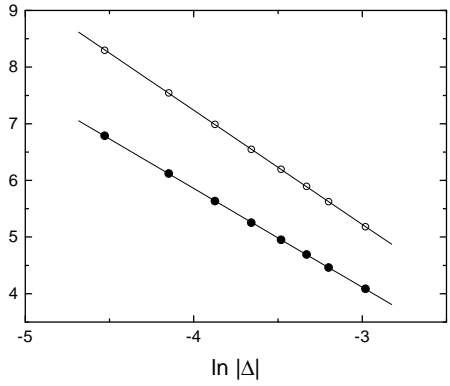}
\caption{
Log-log plots of $R_s$ and $R_s / |\ln |\Delta||$ versus $|\Delta|$. 
Open circles are for $R_s$ and filled circles for $R_s / |\ln |\Delta||$.
The solid lines are of slope -2.00 and -1.75, respectively.
}
\label{fig4}
\end{figure}

\begin{figure}
\centering
\includegraphics[height=8cm,angle=270]{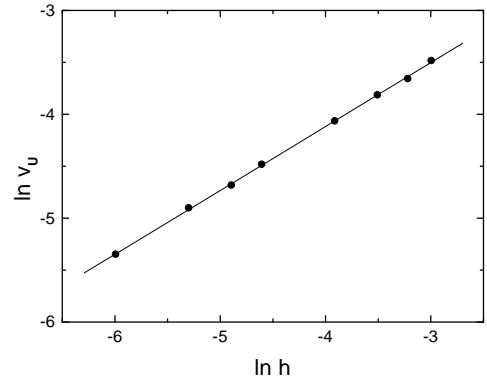}
\caption{
Log-log plots of $v_U$ versus $h$ at $p=p_c^0=0.125$. 
The solid line is of slope 0.62.
}
\label{fig5}
\end{figure}

\end{multicols}

\begin{references}
\bibitem{explain2} This is the conventional definition of the dynamic
             exponent $z$. We do not take the traditional usage of $z$ as the exponent of 
             the mean-square distance of spreading at the absorbing transition. This
             spreading exponent can be expressed as $2/z$.
\bibitem{explain0} The values of the critical probabilities were
measured more accurately in the defect-type Monte Carlo simulations.
\end{references}
\end{document}